\documentclass[sn-mathphys,Numbered]{sn-jnl}

\usepackage{graphicx}%
\usepackage{multirow}%
\usepackage{amsmath,amssymb,amsfonts}%
\usepackage{amsthm}%
\usepackage{mathrsfs}%
\usepackage[title]{appendix}%
\usepackage{xcolor}%
\usepackage{textcomp}%
\usepackage{manyfoot}%
\usepackage{booktabs}%
\usepackage{algorithm}%
\usepackage{algorithmicx}%
\usepackage{algpseudocode}%
\usepackage{listings}%

\theoremstyle{thmstyleone}%
\theoremstyle{thmstyletwo}%

\theoremstyle{thmstylethree}%

\def \doiurl#1{\url{https://doi.org/#1}}

\begin{document}

\title[A Lagrangian path integral approach to the qubit]{A Lagrangian path integral approach to the qubit}


\author*[1,2]{\fnm{Alberto} \sur{Ibort}}\email{albertoi@math.uc3m.es}

\author[1]{\fnm{Mar\'{\i}}a \sur{Jim\'enez-V\'azquez}}\email{maria.j.vazquez@alumnos.uc3m.es}
\equalcont{These authors contributed equally to this work.}

\affil*[1]{\orgdiv{Department of Mathenatics}, \orgname{Universidad Carlos III de Madrid}, \orgaddress{\street{Avda. de la Universidad 30}, \city{Legan\'es}, \postcode{28911}, \state{Madrid}, \country{Spain}}}

\affil[2]{\orgdiv{ICMAT}, \orgname{Instituto de Ciencias Matem\'{a}ticas (CSIC-UAM-UC3M-UCM)}, \orgaddress{\street{C. Nicol\'as Cabrera, 13--15}, \city{Fuencarral-El Pardo}, \postcode{28049}, \state{Madrid}, \country{Spain}}}


\abstract{
A Lagrangian description of the qubit based on a generalization of Schwinger's picture of Quantum Mechanics using the notion of groupoids  is presented.    In this formalism
a Feynman-like computation of its probability amplitudes is done.   The Lagrangian is interpreted as a function on the groupoid describing the quantum system.   Such Lagrangian determines a self-adjoint element on its associated algebra.    Feynman's paths are replaced by histories on the groupoid which form themselves a groupoid. A simple method to compute the sum over all histories is discussed.   The unitarity of the propagator obtained in this way imposes quantization conditions on the Lagrangian of the theory.   Some particular instances of them are discussed in detail.
}




\maketitle

\tableofcontents

\section{Introduction:  The Lagrangian and Quantum Mechanics}\label{sec:intro}

One of the most significant conceptual developments of Quantum Mechanics took place when, following Dirac's insight \cite{Di33}, both R. Feynman and J. Schwinger offered their own separate interpretations for the role of the Lagrangian in the foundations of the theory.  Feynman \cite{Fe48,Fe05} by means of his celebrated path integral description of probability amplitudes and Schwinger \cite{Sc91,Sc01}, by invoking a quantum variational principle from which probability amplitudes could be derived.  It is worthwhile to point out that both approaches are profoundly different, the most significant difference being the interpretation of the Lagrangian itself.  While in Feynman's approach it is just the classical Lagrangian function $L$ defined on the classical velocity phase space $TQ$, with $Q$ the configuration space of a classical system associated to the quantum system under study, in Schwinger's picture, the Lagrangian $\mathbf{L}$ is an operator-valued function of the basic quantum observables of the theory.  

While both approaches render the same results for a large class of quantum mechanical systems and quantum field theories, more precisely, those that have a ``good'' classical description in terms of standard classical mechanical or classical field notions, it is not obvious at all how to use the fundamental ideas of both theories when dealing with simple quantum mechanical system without a classical counterpart, that is, systems with no obvious classical description like, for instance, the qubit.   In fact, when describing the quantum mechanical properties of the qubit we must rely on the standard description of quantum mechanical systems based on Hilbert spaces and linear operators as established by von Neumann, Dirac, etc.    There is no path integral description of the system consistent with Feynman's principle because, to begin with, there is no classical Lagrangian associated to it.   

It must be pointed out now that there is a Schwinger's-like description of the qubit by using Schwinger's picture of quantum mechanics based on the algebraic structure provided by his algebra of selective measurementes \cite{Sc59}.  In the wider perspective offered by the groupoids description of quantum mechanical systems recently put up by Ciaglia \textit{et al}  that clarifies and extends Schwinger's picture of quantum mechanics, the description of the qubit is carried on without pain (see, for instance, examples in \cite{Ci19,Ci19a,Ci19b,Ci19c,Ci24}), however such description is still unsatisfactory because again, there is no Lagrangian description of it.    In any case, even if the groupoids description of quantum mechanics, that we often refer as ``Schwinger's picture of quantum mechanics'', provides a simple, natural, framework to describe quantum mechanical systems lacking a classical Lagrangian description,  there is no hint in Schwinger's original work how to apply his own variational principle to such systems because, at it was pointed out before, no obvious Lagrangian operator is associated to such systems.

The main contribution of the present work is to offer a new understanding of the dynamics of simple quantum mechanical systems like the qubit that do not possess an obvious classical counterpart by combining ideas from Feynman's path integral approach and Schwinger's picture of quantum mechanics.    In fact, in what follows a Feynman-like path integral description of the qubit will be presented based on the groupoids picture of quantum mechanics inspired on Schwinger's seminal contributions.   In doing that the role of the Lagrangian for such systems will be elucidated and it will be identified with a function $\ell$ defined on the groupoid $\Gamma$ used to describe the system under study satisfying appropriate conditions (see \S \ref{sec:lagrangian} for details).    Then, the groupoid itself will play the role of the  classical velocity tangent space $TQ$ of the theory and the function $\ell$ could be interpreted both as a classical function but also as an element in the von Neumann algebra of the groupoid, i.e., as an element in the algebra of observables of the theory (see \S \ref{sec:observables}).   In this way a natural relation between Schwinger's operatorial interpretation of the Lagrangian and Feynman's classical interpretations is obtained.   

In this context the groupoid $\Gamma$  itself carries all quantum mechanical properties of the system while the  classical velocity phase space $TQ$ of the theory, whenever it exists, provides an infinitesimal description of its quantum properties.  For systems, like the instances considered here the groupoid used to describe them has not an infinitesimal description and the standard Feynman's description fails, while it is still possible to address a Feynman-like path integral description by extending the notion of Feynman's path integral  to the groupoid itself.     The main objective of this work is to show explicitly how this can be done for the qubit. 

The paper will be organized as follows.   In Sect. \ref{sec:qubit} the basic ideas of the groupoidal/Schwinger picture of quantum mechanics will be  succinctly reviewed and the particularly simple example of the qubit will be described in detail.   The essence of this new description lies in associating to each experimental setting used to describe a quantum system, a groupoid (see, for instance, \cite{Ib19}, for a gentle introduction to the theory of groupoids), whose mathematical structure captures the properties of the outcomes and transitions of quantum systems.

The next step would be to apply these  concepts in calculating transition amplitudes. To achieve this, Sect. \ref{sec:feynman} will be devoted to discuss the extension of Feynman's path integral principle to the groupoids setting. Following that, the natural emergence of crucial concepts, such as the notion of \textit{history} (that will extend Feynman's \textit{paths} and will introduce the \textit{groupoid of histories} of the system), or the notion of \textit{DFS states}, that will generalise Feynman's principle assigning a complex number to each history, will be presented concluding with the derivation of a new general expression for probability amplitudes in this formalism (\S \ref{sec:feynman_path}).

At last, \S \ref{sec:qubit_amplitude}, the previous findings are applied specifically to a two-level quantum system known as a qubit. The probability amplitude matrix for this system is derived and it comes out as a surprise that imposing the unitarity of the propagator of the theory implies that the values of the parameters of the Lagrangian of the system are quantized.  Explicit expressions are derived and a preliminary discussion will be offered.  The paper concludes with a succinct summary of the main conclusions derived throughout its elaboration in \S \ref{sec:conclusions}.\\


\section{Quantum systems and groupoids: the qubit}\label{sec:qubit}

\subsection{The kinematical description of quantum systems: outcomes and transitions: the qubit and other simple examples}
The groupoids based picture of Quantum Mechanics is a natural extension of Schwinger's algebra of selective measurements.  Its fundamental assumption is that to any experimental setting used to describe a quantum system we associate both the collection of all possible outcomes and all possible transitions  that can be described by means of the observations and measurements performed using the devices provided by such setting.   

The outcomes of the system will be denoted by $a,b,x,y, \ldots$ and they will form a set $\Omega$ called the space of outcomes of our system.  On the other hand, the transitions of the system will be denoted by $\alpha \colon a \to b$, meaning by that that the outcome of the system immediately before the transition takes place would have been $a$, and right after the transition it would be $b$.    The total collection of transitions will be denoted by $\Gamma$.   Note that there are two natural maps, $s,t \colon \Gamma \to \Omega$, called the source and the target, that assigns to the transition $\alpha \colon a \to b$, its source $a$, $s(\alpha) = a$, and its target $b$, $t(\alpha) = b$, respectively.      

The occurrence of two transitions $\alpha \colon a \to b$, $\beta \colon b \to c$, one right after the other, defines a natural composition law of transitions $\beta \circ \alpha \colon a \to c$.    Note that two transitions $\alpha$ and $\beta$ can be composed only if $s(\beta) = t (\alpha)$, in which case it will be assumed that the composed transition $\beta \circ \alpha$ will exists and the transitions $\beta, \alpha$ will be said to be composable.    The collection of pairs of transitions $(\beta, \alpha) \in \Gamma \times \Gamma$, that can be composed will be denoted $\Gamma^{(2)}$ and the composition law $\circ$ is defined in the set $\Gamma^{(2)}$.    It will be assumed that the partial composition law is associative, meaning by that that $(\gamma \circ \beta) \circ \alpha = \gamma \circ (\beta \circ \alpha)$, provided that $(\gamma, \beta)$ and $(\beta, \alpha)$ are composable.    It is also natural to assume that there are transitions that do not affect the observations of the system, in other words, for any outcome $a \in \Omega$, it will be assumed the existence of a transition $1_a \colon a \to a$, such that $\alpha \circ 1_a = \alpha$, and $1_b \circ \alpha = \alpha$, for any $\alpha \colon a \to b$.   The transitions $1_a$ will be called units and they are in one-to-one correspondence with the outcomes $a$ of the system.

The most important nontrivial property that will be assumed to hold for the observed transitions of a quantum physical system is the existence of inverses, that is, given $\alpha \colon a \to b$, we will assume that there exists the transition $\alpha^{-1} \colon b \to a$, such that $\alpha^{-1} \circ \alpha = 1_a$, and $\alpha \circ \alpha^{-1}  = 1_b$.   The physical justification for such assumption lies in Feynman's microreversibility principle stated forcefully in \cite[p. 3]{Fe05}: ``\textit{The fundamental (microscopic) phenomena in nature are symmetrical with respect to the interchange of past and future}''.    Of course, it could happen that the actual experimental setting used in performing the experiments upon which we will build our description of the system will lack the capability to test all the previously stated properties.     In this sense the family of properties for the composition of transitions should be understood as an idealization of the actual experimental settings where all relevant experiments could be carried on.

The structure determined by the collection of all transitions $\Gamma$ together with the partial composition law $\circ$ satisfying the properties above are call an algebraic groupoid.  Sometimes, to emphasize the role of the source and target maps as well as the space of outcomes, we will denote the groupoid $\Gamma$ as $\Gamma \rightrightarrows \Omega$.      We can summarize the previous discussion by saying that the description of a given quantum system provided by an adequate experimental setting determines a groupoid $\Gamma \rightrightarrows \Omega$ (see \cite{Ci19,Ci19a,Ci19b,Ci19c,Ci20a,Ci20b,Ci24} for detailed discussions on the fundamental ideas around the notion of groupoids and the description of quantum systems, in particular find in \cite[\S 4]{Ci19a}  the explicit description of the relation of the abstract groupoids picture of quantum mechanics and Schwinger's algebra of selective measurements).

In what follows we will provide the groupoids description of the two families of quantum systems mentioned in the introduction: the qubit (and any finite level quantum systems) and quantum systems associated to classical mechanical systems.

\subsubsection{The Qubit}
The qubit is arguably the simplest non-trivial quantum system, nevertheless it raises all fundamental conceptual issues facing the description of quantum systems.  The system has only two outcomes, denoted in what follows by $+$ and $-$, thus the space of outcomes of the system is $\Omega = \{ + , - \}$, and it has only two non-trivial transitions, those corresponding to the observation of the system changing the outcome $-$ by $+$ and, conversely, $+$ by $-$. 
They will be denoted by $\alpha \colon + \to -$, and $\alpha^{-1} \colon - \to +$.   Together with them there are the units $1_+ \colon + \to +$, $1_-\colon - \to -$, that do not affect the registered outcomes of the system.   Thus the groupoid describing such system is the set $\{ 1_+, 1_-, \alpha, \alpha^{-1}\}$ that will be denoted by $A_2$ (see Fig. \ref{fig:A2}).    The composition law is defined by the relations collected in the table below, Table \ref{fig:G1} (the content of each cell is the composition of the transition in the corresponding column with the transition appearing in the corresponding row, the symbol $\ast$ indicates that the two transitions involved are not composable), for instance $\alpha^{-1}\circ \alpha = 1_+$, $\alpha^{-1}\circ 1_- = \alpha^{-1}$.

\begin{table}[htp]
\begin{tabular}{|c|c|c|c|c|}
\hline     $A_2$        & $1_+$ & $1_-$ & $\alpha$  & $\alpha^{-1}$ \\ \hline
$1_+$         &  $1_+$          & $\ast$         & $\ast$            & $\alpha^{-1}$          \\ \hline
$1_-$         & $\ast$         &  $1_-$          &  $\alpha$          &  $\ast$                 \\ \hline
$\alpha$      &  $\alpha$       &  $\ast$         & $\ast$            & $1_-$                  \\ \hline
$\alpha^{-1}$ & $\ast$         & $\alpha^{-1}$  & $1_+$         & $\ast$   \\       \hline
\end{tabular}
\caption{Multiplication table for the groupoid $A_2$}\label{fig:G1}
\end{table}%

The groupoid $A_2$ can be understood, for instance, as an abstract algebraic description of a Stern-Gerlach apparatus and the corresponding experiments performed upon spin 1/2 particles.  The outcomes $+,-$ will be identified with the spots in the plates obtained when the particles traverse the apparatus, thus identifying $+$ with the spot in the upper part of the $z$-axis of the screen (positive $z$-component of the spin) and $-$ with the spot in the lower part of the axis (negative $z$-component of the spin, $-1/2$).  The transitions $\alpha$, $\alpha^{-1}$ will correspond to the transitions observed when a particle that has emerged with positive (respec., negative) $z$-component spin, traverse again the apparatus and emerges with negative (respec., positive) $z$-component spin.   Diagrammatically, the groupoid $A_2$ can be represented as follows:

\begin{figure}[h]%
\centering
   \resizebox{5cm}{1cm}{\includegraphics{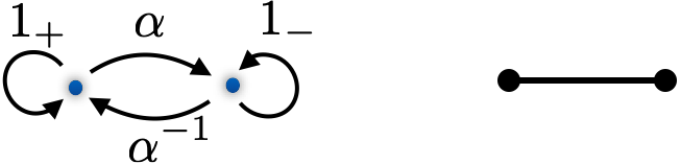}} 
\caption{Left: Diagrammatic representation of the qubit groupoid.  Right: the $A_2$ Dynking diagram.}\label{fig:A2}
\end{figure}

Note that the qubit groupoid $A_2$ is not commutative (as it is shown by the fact that the multiplication table of the groupoid, Table \ref{fig:G1}, is not symmetric). 


\subsection{Observables: the algebra of a groupoid}\label{sec:observables}

The groupoid $A_2$ can be realized in terms of $2\times 2$ matrices.   That is, the assignment:
$$
1_{-} \mapsto \mathbf{1}_- = \left(\begin{array}{ll}
1 & 0 \\
0 & 0
\end{array}\right), \, 1_{+} \mapsto \mathbf{1}_+  = \left(\begin{array}{ll}
0 & 0 \\
0 & 1
\end{array}\right), \, \alpha^{-1} \mapsto A^* = \left(\begin{array}{ll}
0 & 1 \\
0 & 0
\end{array}\right), \, \alpha \mapsto A = \left(\begin{array}{cc}
0 & 0 \\
1 & 0
\end{array}\right) .
$$
preserves the multiplication table of the groupoid as it is easily checked by direct inspection.   Moreover the algebra of matrices generated by the matrices  $\mathbf{1}_\pm, A, A^*$ is the algebra $M_2(\mathbb{C})$ of all $2\times 2$ matrices.
This reflects the fact that to any groupoid $\Gamma$ we can associate an algebra, called the algebra of the groupoid, and that, in the case of the qubit as well as in the case of finite level quantum systems, such algebra is just an algebra of matrices.      If $\Gamma$ is a finite groupoid we will denote such algebra as $\nu (\Gamma)$ and its elements $a \in \nu (\Gamma)$ are just formal linear combinations of the transitions in $\Gamma$, that is, $a = \sum_{\alpha \in \Gamma} a_\alpha \, \alpha$, $a_\alpha \in \mathbb{C}$.   There is a natural associative product defined on $\nu (\Gamma)$ by means of the formula:
$$
a \cdot b = \sum_{(\alpha, \beta) \in \Gamma^{(2)}} \,a_\alpha b_\beta \,\,\, \alpha \circ \beta \, ,
$$ 
provided that $a = \sum_{\alpha \in \Gamma} a_\alpha \, \alpha$, and $b = \sum_{\beta \in \Gamma} b_\beta \, \beta$.     The algebra $\nu (\Gamma)$ is unital and the unit is given by $\mathbf{1} = \sum_{a \in \Omega} 1_a$.   

In the simplest case that our groupoid $\Gamma$ is the groupoid of pairs $P(\Omega_n)$ of a finite set $\Omega_n = \{ x_1, x_2, \ldots, x_n \}$, i.e., $\Gamma = \Omega_n \times \Omega_n$, with composition law $(x_i, x_j) \circ (x_j, x_k) = (x_i,x_k)$, for all $i,j,k = 1, \ldots, n$, the corresponding algebra of the groupoid is identified in a natural way with the algebra of matrices $M_n(\mathbb{C})$, the assignment being that to any element $a = \sum_{i,j= 1}^n a_{ij} \, (x_i,x_j)$, we associate the matrix $A$ whose entries are the coefficients $a_{ij}$ of the element $a \in \nu (\Gamma)$.   This is precisely the form that the algebra of selective measurements took in Schwinger original's presentation, however we should not confuse a representation of the algebra of a groupoid $\nu (\Gamma)$ with the abstract algebra itself, and much less with the groupoid the determines it. 

The algebra of a groupoid carries another natural operation that reflects directly the inverse law of the groupoid that associates to any transition $\alpha$ its inverse $\alpha^{-1}$ and that will be denoted by $\tau$, i.e., $\tau(\alpha ) = \alpha^{-1}$.  It is given by an involution operator $*$ on $\nu (\Gamma)$ that assigns to any element $a = \sum_{\alpha \in \Gamma} a_\alpha \, \alpha$, the element $a^* = \sum_{\alpha \in \Gamma} \bar{a}_\alpha \, \alpha^{-1}$.    Note that the transitions $\alpha$ of the groupoid are elements of the algebra $\nu(\Gamma)$ themselves, those corresponding to elements $a$ whose coefficients are given by $\delta_\alpha$, where $\delta_\alpha$ is the function on $\Gamma$ defined as $\delta_\alpha (\beta)= 1$ if $\beta = \alpha$, and 0 otherwise.   We will denote such element in $\nu (\Gamma)$ with the same symbol  if there is no risk of confusion.  Note that with these conventions $\alpha^* = \alpha^{-1}$.      The involution operation $*$  is an antilinear map from $\nu (\Gamma)$ into itself and satisfies $(a^*)^* = a$, and $(ab)^* = b^* a^*$, for all $a,b \in \nu (\Gamma)$.  It is obvious that the involution operation in the previous examples, the algebra of the qubit and the algebra of the groupoid of pairs of a finite set, correspond to the standard adjoint operator $A\mapsto A^\dagger$ in the algebra of matrices representing the algebras of both groupoids (see, for instance, \cite{Ci19b,Ib19} for detailed descriptions of the algebras of groupoids).

There is a natural representation of the algebras of finite groupoids, called the fundamental representation, that can be described simply as the natural representation $\pi_0$ on the Hilbert space generated by the outcomes of the groupoid $\Gamma \rightrightarrows \Omega$, that is, we construct the Hilbert space $\mathcal{H}_0$ generated by vectors $| a \rangle$, $a \in \Omega$, with the inner product that makes them an orthonormal basis, i.e., $\langle b \mid a \rangle = \delta_{ab}$.    Then, we define $\pi_0 (\alpha) |a\rangle = |b \rangle$, provided $\alpha \colon a \to b$.     By linearity such representation associates a bounded operator $A = \pi_0(a)$ to any element $a \in \nu (\Gamma)$ and, as it is easy to check $\pi_0(a^*) = \pi_0(a)^\dagger$.   

In the case of the qubit groupoid $A_2$, the Hilbert space $\mathcal{H}_0$ supporting the fundamental representation will be generated by two orthonormal vectors $|+ \rangle$, $|-\rangle$, or identifying $\mathcal{H}_0$ with $\mathbb{C}^2$, we may say that the outcomes $+,-$ or the system are represented by the vectors:
$$
-\mapsto | - \rangle = \left(\begin{array}{l}
1 \\
0
\end{array}\right) \, , \qquad  +\mapsto |+ \rangle= \left(\begin{array}{l}
0 \\
1
\end{array}\right),
$$
and then, any element of the algebra $a \in \nu (A_2)$ will be represented by the matrix:
$$
\pi_0(a)=\left(\begin{array}{cc}
a_{-} & a_{\alpha^{-1}} \\
a_\alpha & a_{+}
\end{array}\right).
$$
Thus, the representations that were defined previously for both the algebra of the qubit groupoid $A_2$ and the groupoid of pairs  $P(\Omega_n)$ of any finite set are just the fundamental representation of them (see, for instance, \cite{Ib19} for a detailed exposition of the theory of representations of finite groupoids).     

The fundamental representation allows to introduce a natural norm on the algebra $\nu (\Gamma)$ as $|| a || := || \pi_0(a) ||_2$, where the norm in the right is the Hilbert space norm in $\mathcal{H}_0$.   The norm $|| \cdot ||$ thus defined satisfies $|| a^* a || = || a ||^2$, and it determines a von Neumann algebra structure on the algebra $\nu (\Gamma)$.  Then, the algebra $\nu(\Gamma)$ will be called the von Neumann algebra of the groupoid $\Gamma$.  For finite groupoids $\Gamma$, the von Neumann algebra $\nu (\Gamma )$ is isomorphic to a direct sum of Type $I_n$ factors, that is, algebras isomorphic to $M_n (\mathbb{C})$.   

Observables of the quantum system defined by the groupoid $\Gamma$ are the self-adjoint elements $a = a^*$, in the von Neumann algebra $\nu (\Gamma)$ of the groupoid, in other words, $a = \sum_{\alpha \in \Gamma} a_\alpha \, \alpha$ is self-adjoint if $\bar{a}_\alpha = a_{\alpha^{-1}}$.

The previous notions become more involved when the groupoids used to describe a quantum system are not finite or carry additional structures like, for instance, a measure structure (see \cite{Ci24} and references therein).    For instance, the groupoidal description of a system associated to a classical mechanical system with configuration space $Q$ is carried on by considering the groupoid of pairs $P(Q) = Q\times Q$ of $Q$.   Typically $Q$ is a Riemannian manifold carrying a metric $g$, for instance $\mathbb{R}^3$ with the standard Euclidean metric in the case of a Newtonian particle.    The transitions $(q_2,q_1) \in P(Q)$, with $q_1, q_2 \in Q$ in the groupoid are interpreted as experiments where the particle is detected first with configuration $q_1$ and later on is detected with configuration $q_2$.    Such groupoid carries a natural smooth structure and a measure structure (determined by the volume defined by the metric $g$).    In such case it can be shown \cite{Ci19b,Ci24} that the von Neumann algebra of $P(Q)$ is the von Neumann algebra of all bounded operators in the Hilbert space $L^2(Q)$ of square integrable functions on $Q$ (which is a factor of Type $I_\infty$).    Much more complicated von Neumann algebras can emerge as von Neumann algebras of groupoids, for instance if we consider the qubit groupoid $A_2$ carrying a probability measure $p = (p_+,p_-)$, $p_\pm \geq 0$, $p_+ + p_- = 1$, on the space of outcomes $\{ +, -\}$ of the system, the von Neumann algebra of an infinite chain of such gruopoids, that is $A_2^\infty$ (interpreted as a spin chain), can be shown to be a Type $III_\lambda$ factor, with $0 < \lambda = p_+ < 1/2$ \cite{Ci23}.   The measure $p$ on the space of outcomes of the qubit could be understood in physical terms as an intrinsic bias of the system, that is, as a background mixing in the system described by the probability vector $p$.  This classical background probability will appear again in the actual path integral computation of probability amplitudes for the qubit system (see Sect. \ref{sec:qubit_amplitude}).


\subsection{States and dynamics}\label{sec:states}

This succinct review of fundamental notions in the groupoids description of quantum systems used to understand the Lagrangian dynamical description of the qubit will end up describing their Hamiltonian dynamics.   Given a groupoid $\Gamma$ describing a quantum system, a Hamiltonian description of their dynamics is easily introduced either by using, for instance, the fundamental representation described before or, more intrinsically, by considering a Hamiltonian observable $h$ in the groupoid, that is the Hamiltonian of the system will be a self-adjoint element $h = h^*$ in the von Neumann algebra $\nu (\Gamma)$ of the groupoid $\Gamma$.    A self-adjoint element defines a one parameter group of automorphisms $\phi_t$ of $\nu (\Gamma)$ by means of $\phi_t = \exp (-\frac{i}{\hbar}th)$, or, written as an equation in the algebra of the groupoid, the evolution of the system is given by the Heisenberg-like equation of motion:
\begin{equation}\label{eq:heisenberg}
\frac{da}{dt} = i \hbar  [a,h] \, \qquad a \in \nu (\Gamma) \, ,
\end{equation}
where $[a,h]$ denotes the commutator $a\cdot h - h \cdot a$ of $a$ and $h$.   Note that in terms of the operators $A = \pi_0(a)$, $H = \pi_0(h)$, associated to the elements $a,h$ by means of the fundamental representation $\pi_0$, Eq. (\ref{eq:heisenberg}), takes the standard form:
$$
\frac{dA}{dt} = i \hbar [A,H]. \
$$
The observables of the groupoid $\Gamma$ can also be thought as functions on the groupoid satisfying a reality condition.  Indeed, if $a = a^*$ is an observable, we may think of it as defining a function $f_a \colon \Gamma \to \mathbb{C}$, $f_a (\alpha) = a_\alpha$, where $a = \sum_{\alpha} a_\alpha \, \alpha$.  The function $f_a$ satisfies $f_a(\alpha^{-1}) = \overline{f_a(\alpha)}$.  This dual interpretation of observables, as self-adjoint elements in the von Neumann algebra of the groupoid, and as numerical valued functions on the groupoid satisfying a reality condition, will be instrumental in the definition of the Lagrangian of the qubit and its physical interpretation (see, Eq. (\ref{eq:I1}),  Sect. \ref{sec:lagrangian}).

Finally, we will devote a few words to talk about states in the groupoidal picture of quantum mechanics (see, for instance, \cite{Ci19c,Ci20a,Ci20b,Ci20c,Ci21b,Ci24}, for more details).   Given a quantum system described by the groupoid $\Gamma$, the quantum states of the system are described by normalized positive functionals $\rho$ on the von Neumann algebra $\nu (\Gamma)$ of the groupoid, that is $\rho \colon \nu(\Gamma) \to \mathbb{C}$, such that $\rho$ is linear, $\rho (\mathbf{1}) = 1$, and $\rho (a^* a ) \geq 0$ for all $a\in \nu (\Gamma)$.     Given a representation of the algebra, vectors in the supporting Hilbert space of the representation can be understood as states of the system.   For instance, if we consider the fundamental representation $\pi_0$ of the groupoid, then any vector $| \psi \rangle \in \mathcal{H}_0$, defines a state $\rho_\psi$ by means of:
$$
\rho_\psi (a) = \frac{\langle \psi | \pi_0(a) | \psi \rangle}{\langle \psi | \psi \rangle} \, , \qquad \forall a \in \nu (\Gamma) \, .
$$
The state $\rho_\psi$ thus constructed is an example of a pure state.   Of course, density operators $W$ in $\mathcal{H}_0$ gives rise to states of the system by means of $\rho_W (a ) = \mathrm{Tr\,} (W \pi_0(a))$.     The evolution of the system can also be stated in terms of states but we will not need this approach in this work.


\section{The path integral description of quantum systems in the groupoids picture}\label{sec:feynman}

\subsection{From Feynman's paths to histories: the qubit again}

\subsubsection{Feynman's path integral picture of quantum mechanics}  
Feynman's picture of quantum mechanics provides an explicit expression for the probability amplitude $\langle x_1, t_1; x_0,t_0 \rangle$ for quantum mechanical systems with have a classical Lagrangian description.
Indeed, Feynman used the principle of superposition to calculate such probability amplitude by establishing that each possible path $\gamma = \gamma (t)$, $\gamma (t_0) = x_0$, $\gamma (t_1) = x_1$, relating both outcomes contribute as:
$$
\varphi (\gamma) = C(\gamma) e^{\frac{i}{\hbar} S(\gamma)} \, ,
$$
where $S(\gamma) = \int_{t_0}^{t_1} L(\gamma(t), \dot{\gamma}(t)) dt$, denotes the classical action of the system along the path $\gamma$, defined by the classical Lagrangian of the system $L$, and  $C(\gamma)$ is a normalization constant that will be fixed accordingly with the system to be considered.  A path $\gamma \colon [t_0,t_1] \to Q$, such that $\gamma (t_0)= x_0$, and $\gamma(t_1) = x_1$, will be denoted as $\gamma \colon (x_0,t_0) \to (x_1,t_1)$.  Then Feynman's principle can be written as:
\begin{equation}
\langle x_1, t_1; x_0,t_0 \rangle = \sum_{\gamma:(x_0,t_0)\rightarrow(x_1,t_1)} \varphi (\gamma) \, ,
\end{equation}
or, using an integral notation, as:
\begin{equation}\label{eq:feynman_path}
\langle x_1, t_1; x_0,t_0 \rangle = \int_{\gamma:(x_0,t_0)\rightarrow(x_1,t_1)} \mathcal{D}\gamma \,\,\, C(\gamma) e^{\frac{i}{\hbar} S(\gamma)} \, ,
\end{equation}
where the symbol ``$\int \mathcal{D}\gamma\,$'' indicates a (non-specified yet) way of computing such sum over the space of all paths joining $x_0$ and $x_1$.     
  
 Certainly, the way to perform the summation over all possible paths is not trivial as it was acknowledged by Feynman himself \cite[Sect. 2.4]{Fe05}: \textit{``The number of paths is a high order of infinity, and it is not evident what measure is to be given to the space of paths. It is our purpose in this section to give such a mathematical definition. This definition will be found rather cumbersome for actual calculation [...]  As for this section, it is hoped that the mathematical difficulty or rather inelegance, will not distract the reader from the physical content of ideas.''}      It is the purpose of the coming sections to provide an extension of Feynman's principle that will allow us to compute the transition amplitudes for the qubit, first by providing the adequate extension of the notion of path when the system we are dealing with is not defined on a classical space-time and, secondly, by introducing a consistent way to compute the sum over all such generalized paths.  We hope, together with Feynman, that the mathematical difficulty or inelegance of the methods presented here will not distract the reader from the physical content of the ideas.

One way to describe mathematically the symbol $\int \mathscr{D}\gamma$ resorts to the standard description of quantum mechanics in terms of the position $q$ and momentum $p$ operators in a Hilbert space $\mathcal{H}$.  Then the probability amplitude of a particle propagating from the point $x_i$ to $x_f$ is:
\begin{equation*}\label{completitud}
\langle x_f,t_f; x_i,t_i \rangle =\langle x_f| \;e^{-\frac{i}{\hbar}H(t_f-t_i)}\; | x_i \rangle \, .
\end{equation*}
The amplitude is completely determined by the unitary operator $\exp{-\frac{i}{\hbar} Ht}$ with $H$ the hamiltonian of the system. Then, the time $t_f - t_i$ in which the transition takes place is discretized considering $N$ subintervals of length $\tau = (t_f - t_i)/N$, and after an appropriate use of the completeness and orthonormality relations of $q$ and $p$, and taking the continuous limit for time, i.e., $N\rightarrow\infty$, we get:
$$
\langle x_f,t_f; x_i,t_i \rangle =\langle x_f| \;e^{-\frac{i}{\hbar}H(t_f-t_i)}\; | x_i \rangle =  \int_{\gamma:(x_0,t_0)\rightarrow(x_1,t_1)} \mathcal{D}\gamma \,\,\, C(\gamma) e^{\frac{i}{\hbar} S(\gamma)} \, ,
$$
where:
\begin{equation}\label{eq:measure_paths}
\int_{\gamma: (x_i,t_i) \rightarrow (x_f,t_f)} \mathcal{D} \gamma =``\lim _{N \rightarrow \infty} "\left(-\frac{2 \pi m i}{\Delta t}\right)^{N / 2} \int  dq_{N-1} dq_{N-2} \cdots dq_2  dq_1 \, ,
\end{equation}
when considering a point particle of mass $m$ with classical Lagrangian function $L = \frac{1}{2}m v^2$, and suitable choices for the normalization constants $C(\gamma)$. 

\subsubsection{Beyond Feynman's paths: histories} \label{sec:histories}
Unfortunately, the previous ideas cannot be implemented straightforwardly in the case of the qubit because of the lack of the ingredients used in the derivation of Feynman's formula, \textit{cfr.} Eq. (\ref{eq:feynman_path}), that is, a configuration space $Q$ and a Lagrangian $L$, however the groupoid approach allows to dispose with these difficulties by extending Feynman's paths to the groupoids setting and introducing a new notion of Lagrangian that will be suitable for our purposes.

The notion of paths used in Feynman's path integral formula extends easily to the setting of groupoids.  If $\Gamma \rightrightarrows \Omega$ is the groupoid describing our quantum system, we will substitute the classical configuration space $Q$ by the space of outcomes $\Omega$ of the groupoid, thus a path would be just a map $\gamma \colon [t_i,t_f] \to \Omega$.   However now, contrary to what happens in the case of a classical system whose groupoid is the groupoid of pairs $Q \times Q$ and where there is only one transition taking $x$ into $y$, the possible transitions from $a$ to $b$ are given by all possible transitions $\alpha \colon a \to b$ in the groupoid $\Gamma$.  Thus if the map $t \mapsto (\gamma(t), x_i) \in Q\times Q$, was the map defined by the path $\gamma \colon (x_i,t_i) \to (x_f,t_f)$ on $P(Q) = Q \times Q$, now, the natural notion of ``path'' with values in the groupoid $\Gamma$, will be a map $w \colon [t_i,t_f] \to \Gamma$, that must start at some outcome $a_0$, that is $w(t_i) = 1_{a_i}$, and that for time $t_f$, it ends up at another outcome $a_f$, that is, $t(w(t_f)) = a_f$.   The map $\gamma(s)$ obtained by projecting $w(s)$ to $\Omega$ by using the target of the transitions $w(s)$, that is, $\gamma (s) = t(w(s))$, recovers the notion of Feynman's path on the space of outcomes $\Omega$.   

We will call \textit{histories} this generalization of the notion of Feynman's paths,  Thus a history $w$ with endpoints $a_i$, $a_f$, at times $t_0$, $t_1$, respectively, is a parametrized curve $w(s)$ on $\Gamma$, starting at $1_{a_0}$ and with final outcome $a_1$.   Histories, as Feynman's paths will be denoted by $w \colon (a_i,t_i) \to (a_f,t_f)$.   There are two natural maps, denoted again by $s,t$, that assign to the history $w\colon (a_i,t_i) \to (a_f,t_f)$,  its ``origin'' $(a_i,t_i)$, and its ``end'' $(a_f,t_f)$, that is, $s(w) = (a_i,t_i)$, and $t(w) = (a_f, t_f)$.

Histories can be composed in a natural way provided that the end of the first agrees with the origin of the second, that is, if $w_1 \colon [t_0,t_1] \to \Gamma$, and $w_2 \colon [t_1,t_2] \to \Gamma$, are the maps describing two histories $w_1$, $w_2$, with endpoints $(a_0,t_0), (a_1,t_1)$ and $(a_1,t_1), (a_2, t_2)$, respectively, then the composition $w_2 \circ w_1$ is a history on the time interval $[t_0, t_2]$, whose associated map is given by:
$$
w_2 \circ w_1 (s) = \left\{ \begin{array}{c} w_1(s) \, , \qquad \mathrm{if\,} t_0 \leq s \leq t_1 \\
w_2(s) \circ w_1(t_1) \, , \quad t_1 \leq s \leq t_2  \end{array} \quad \right.  \, .
$$
It is easy to check that the composition law of histories thus defined is associative.  There are also units for this composition.  Certainly, consider the history $w \colon [t_0,t_1] \to \Gamma$, then the history $\mathbf{1}_{(a_0,t_0)} \colon [t_0,t_0] \to \Gamma$, $\mathbf{1}_{(a_0,t_0}(t_0) = 1_{a_0}$, satisfies that $w \circ \mathbf{1}_{(a_0,t_0} = w$, and $\mathbf{1}_{(a_1,t_1}\circ w = w$.    Thus the space of all histories $w$ is a category with space of objects pairs $(a,t) \in\Omega \times \mathbb{R}$.    We will denote such category $\mathscr{C}^+(\Gamma)$.      

\begin{figure}[h]
  \centering
    \resizebox{11cm}{7cm}{\includegraphics{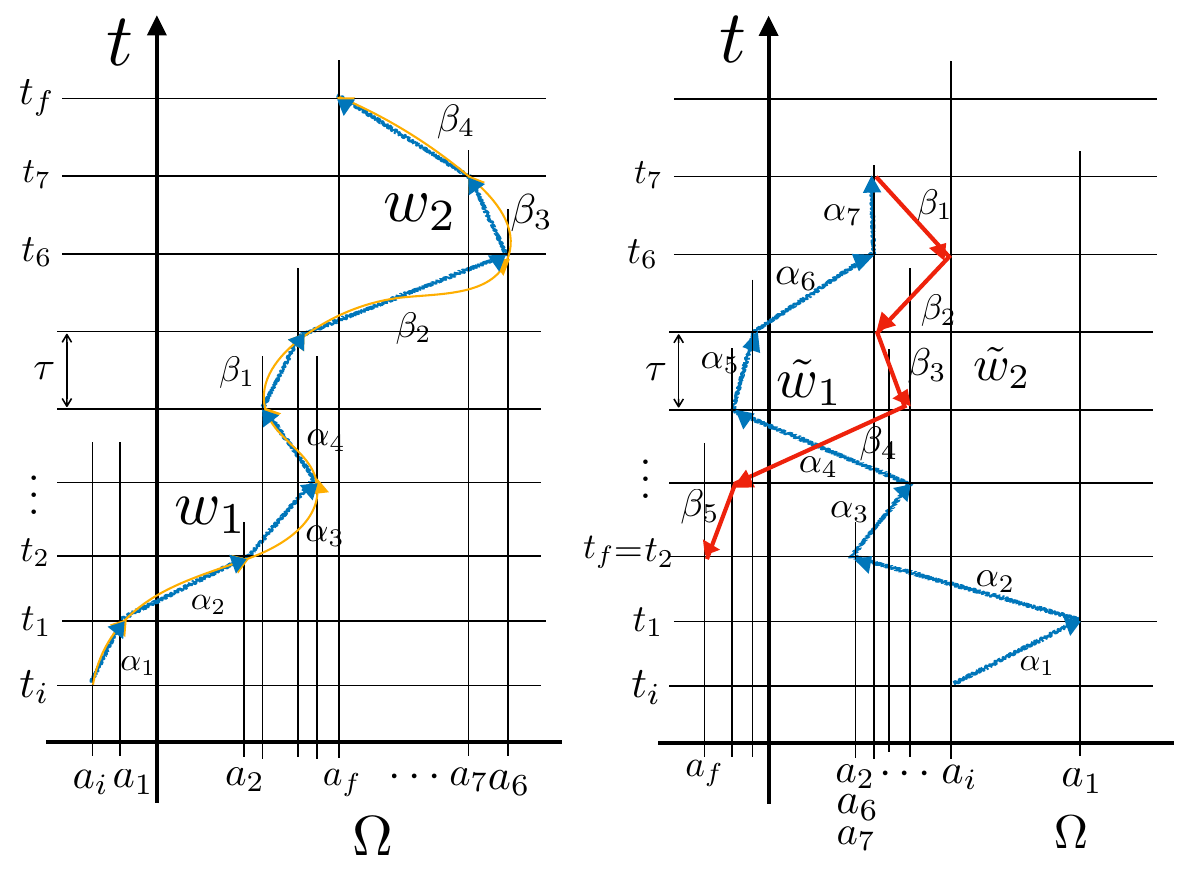}} 
    \caption{Diagrams representing histories on a groupoid.  The vertical axis is the time axis and the horizontal axis represents the space of outcomes $\Omega$ of the system.  On the left two composable histories $w_1, w_2$ are displayed (orange line).  In blue a discrete approximation to $w_1, w_2$ are shown as well as their steps $w_1: \alpha_1, \alpha_2,\alpha_3, \alpha_4$, and $w_2 : \beta_1, \beta_2, \beta_3, \beta_4$.  On the right hand side a future-oriented discrete history $\tilde{w}_1$ (in blue), and a past-oriented history $\tilde{w}_2$ (in red) that can be composed with $\tilde{w}_1$ are shown.}
  \label{fig:histories}
\end{figure}

 As in the derivation of Feynmann's path integral formula, it will be convenient to consider discrete histories, that is, fixed $t_i$, $t_f$, we will partition the interval $[t_i,t_f]$ in $N$ subintervals of length $\tau = (t_f - t_i) /N$, each subinterval of the form $[t_k, t_{k+1}]$,  $t_k = t_i + k\tau$, $k = 0,1,\ldots, N$, thus $t_0 = t_i$, $t_N =t_f$.  Then, we approximate a given history $w$ by the piecewise constant history $\tilde{w} \colon [t_i,t_f] \to \Gamma$, that on each subinterval $[t_k,t_{k+1})$ is constant with value $w_k = w(t_k)$.   Note that $\tilde{w}(t_N) = w(t_f)$.  Hence we can describe the history $\tilde{w}$ by providing the sequence of steps $\alpha_k = \tilde{w}_k \circ \tilde{w}_{k-1}$, $k = 1,2,\ldots , N$ (see Fig. \ref{fig:histories}).  
Note that the sequence: 
 \begin{equation}\label{eq:history_steps}
 \tilde{w} : \, \alpha_1, \alpha_{2}, \ldots, \alpha_{N-1}, \alpha_N \, ,
 \end{equation}
 is such that $\tilde{w}(t_1) = \alpha_1$, $\tilde{w}(t_2) = \alpha_2 \circ \alpha_1$, $\tilde{w}(t_3) = \alpha_3 \circ \alpha_2 \circ \alpha_1$, or, in general:
 \begin{equation}\label{eq:steps}
 w_k = \tilde{w}(t_k) = \alpha_k \circ \ldots \circ  \alpha_1 \, , \qquad \forall k = 1, \ldots, N \, .
 \end{equation}
 
 We will call the total variation of the history $\tilde{w}$, the transition obtained by the composing of its steps, that is, $w_N = \alpha_N \circ \alpha_{N_1} \circ \ldots \circ \alpha_1$.    
 We can complete this discrete description of histories by normalizing them adding a trivial step $\alpha_0$ at the beginning, which is the transition $1_{a_0}$, at $t = t_0$.   Thus, in this sense the history $\tilde{w}(t)$, is recovered by ``integrating'' the steps $\alpha_k$.   Notice that each step is associated to a specific time $t_k$, so when composing the steps in (\ref{eq:steps}), we are following the track of time.     Now, using this discretized description of them, the composition of two histories becomes transparent.  Indeed, if $\alpha_0, \alpha_1, \alpha_2, \ldots, \alpha_N$, is the sequence of steps describing $w_1$ and $\beta_0, \beta_1, \beta_2, \ldots, \beta_M$ is the sequence of steps describing $w_2$, the list of steps describing $w_2 \circ w_1$, will be $\alpha_0, \alpha_1, \alpha_2, \ldots, \alpha_N, \beta_1, \beta_2, \ldots, \beta_M$, i.e,. the history $w_2 \circ w_1$ is obtained by juxtaposition of the sequences of steps describing each one of them.   
 
 Using the previous picture of composition of histories is easy to guess what is the inverse (with respect to the composition of histories $w_2 \circ w_1$ described above) of a given history.   If $\tilde{w}$ is a history with steps $\alpha_0, \alpha_1, \alpha_2, \ldots, \alpha_N$, then the inverse history $\tilde{w}^{-1}$ will have steps $\alpha_N^{-1}, \alpha_{N-1}^{-1}, \ldots, \alpha_1^{-1}, \alpha_0^{-1}$ (pay attention to the reverse order of the steps).   Certainly, when we compose $\tilde{w}^{-1}\circ \tilde{w}$, the corresponding sequence of steps will be $\alpha_0, \alpha_1, \ldots, \alpha_ N, \alpha_{N}^{-1}, \alpha_{N-1}^{-1}, \ldots, \alpha_1^{-1}, \alpha_0^{-1}$.  But the time (reflected in the subbindex $k$ of the steps), starts running backwards when we arrive to the $N+1$th step, and eventully it comes back to the starting time $t_0$. We will call histories such as $\tilde{w}^{-1}$ before, whose steps run backwards in time, past-oriented, in contraposition to histories like (\ref{eq:history_steps}), that will be called future-oriented (see Fig. \ref{fig:histories} for specific examples of future and past-oriented histories and their composition).   
 
 Moreover, the integrated history $\tilde{w}^{-1}\circ \tilde{w}$ is obtained composing the steps,  and its total variation will be the unit $1_{a_0}$ corresponding to the outcome at time $t_i$, so what we get is that the end of the history becomes $\alpha_0 = 1_{a_0}$.   We will identify such history with the unit $\mathbf{1}_{(a_0,t_0)}$ described before.   Note that if we compose a future-oriented history $\tilde{w}_1$ with another past-oriented history $\tilde{w}_2$, such that $t(\tilde{w}_1) = s(\tilde{w}_2)$, we will obtain another history (see Fig. \ref{fig:histories}).  If the past-oriented history $\tilde{w}_2$ ends at the origin of $\tilde{w}_1$, but if the steps $\beta_k^{-1}$ are different form $\alpha_k^{-1}$, then the composed history $\tilde{w}_2 \circ \tilde{w}_1$ will not be trivial (not a unit).  Such histories with the same initial and final outcomes $(a_i,t_i) = (a_f,t_f)$, but non-trivial total variation will be called loops.    
 
 Now, it is clear that the space of all discrete histories, future and past-oriented, is again a groupoid with respect to the composition of histories introduced above.    Such groupoid structure can be extended naturally to the space of continuous histories, even if we will not need to discuss such details here (see, for instance \cite{Ci24}).     In what follows we will just work with discrete histories, so we will denote them just as $w$, $w'$, etc., omitting the symbol ``$\tilde{\quad}$'' that was intended to mean that we were considering a discrete approximation $\tilde{w}$ for the continuous history $w$.   The groupoid of all (discrete) histories on $\Gamma$, will be denoted as $\mathscr{G}(\Gamma)$ or just $\mathscr{G}$ if there is no risk of confusion.  Notice that the space of objects of $\mathscr{G}$ is $\Omega \times \mathbb{R}$, because we need to specify both, the time and outcome of a history to determine whether or not they are composable, thus the consistent notation for the groupoid of histories will be $\mathscr{G} \rightrightarrows \Omega \times \mathbb{R}$.
 
 Now that we have extended Feynman's paths introducing the notion of histories, we would like to be able to associate to each one of them a number $\mathscr{S}(w)$ that will play the role of the action in Feynman's principle and, eventually, of a Lagrangian $\ell$, that will tell us how to compute $\mathscr{S}(w)$. This will be the task of the following section.

\begin{figure}[h]
  \centering
    \resizebox{12.75cm}{5cm}{\includegraphics{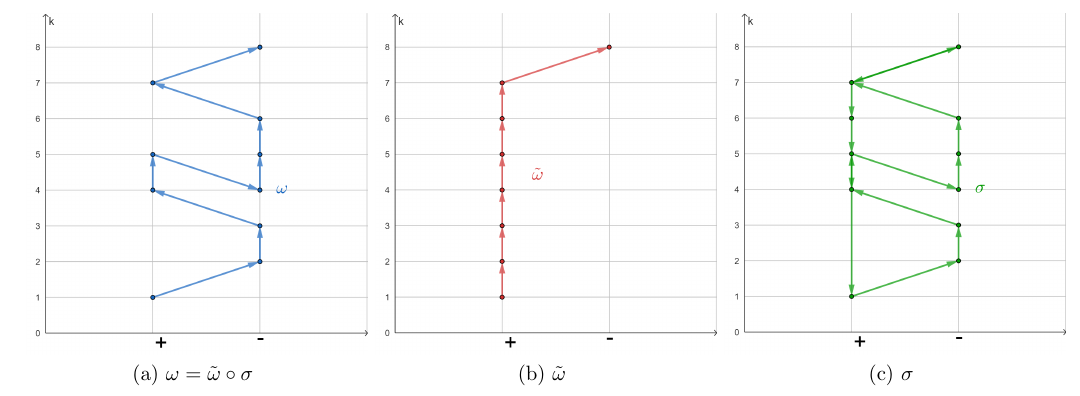}} 
    \caption{Abstract representation of how a history $w$ (in blue) on the qubit groupoid $A_2$ can be decomposed as another history $\tilde{w}$ (in red) with the same origin $(+, \tau )$ and end $(-, 8\tau )$, and a loop $\sigma$ (in green).}
  \label{fig:histories_qubit}
\end{figure}

In the particular instance of the qubit, the quantum system described by the groupoid $A_2$, there are just four transitions $1_\pm, \alpha, \alpha^{-1}$, as possible values of histories.   Hence a future oriented history $w \colon [t_i,t_f] \to A_2$, will have associated steps like, for instance, $1_+, \alpha, 1_-, 1_-, \alpha^{-1}, \ldots$, at times $t_i$, $t_1 = t_i + \tau$, $t_2 = t_i + 2 \tau$, etc.    If there were steps running backwards in time (see Fig. \ref{fig:histories_qubit}), then we can indicate them but underlying them, like $1_-, \alpha^{-1}, 1_+, 1_+, \underline{\alpha}, 1_-, \ldots$.   Note that any history $w$ can be written as a finite composition of future or past oriented histories, thus we will write an arbitrary history $w$ as a composition $w = w_r \circ w_{r-1} \circ \cdots \circ w_1$, where each history $w_l$, $l = 1, \ldots, r$, is oriented.   

The space of histories with fixed beginning and ends, say $(a_0,t_0)$, $(a_1,t_1)$, will be denoted as $\mathscr{G}_{(a_0,t_0)}^{(a_1,t_1)}$, and the space of loops based at $(a_0,t_0)$ will be denoted as $\mathscr{G}(a_0,t_0)$, $\mathscr{G}(a_0)$ if there is no risk of confusion, or just $\mathscr{G}_0$, because the space of loops at $(a_0,t_0)$ is certainly a group, called the isotropy group of $\mathscr{G}$, and all isotropy groups are isomorphic.   Then, it is clear that any history $w\colon (a_i,t_i) \to (a_f,t_f)$, can be written as $w = w_{\mathrm{ref}} \circ \sigma$, where $w_{\mathrm{ref}} \colon (a_i,t_i) \to (a_f,t_f)$ is a given fixed ``reference'' history and $\sigma \in \mathscr{G}(a_0)$ is a loop.  In fact, $\sigma = w_{\mathrm{ref}} ^{-1}\circ w \colon (a_0,t_0) \to (a_0,t_0)$.  It is important to point out that contrary to the notion of paths in Feynman's path integral quantum mechanics, histories are both past and future-oriented. This fact lies behind the previous decomposition of any history as a reference history times a loop and will be instrumental for computing amplitudes as it will be discussed at length in Sect. \ref{sec:feynman_path}.

\subsection{The Lagrangian and the qubit}\label{sec:lagrangian}

Once the notion of path has been extended to the groupoidal setting and we have identified histories as the proper notion, we will need to introduce the dynamics of the system by means of an action.     We will introduce the notion of action of a history abstracting the properties of classical actions, that is, an action will be an assignment $w \to \mathscr{S}(w)$, satisfying:
\begin{eqnarray}
\mathscr{S}(w \circ w') &=& \mathscr{S}(w) + \mathscr{S}(w') \,,  \label{eq:S1} \\
\mathscr{S}(w^{-1}) &=& - \overline{\mathscr{S}(w)} \, ,  \label{eq:S2}
\end{eqnarray}
for any $w$ a history on the groupoid $\Gamma$.      Note that contrary to the situation in classical mechanics we are not restricting the action $\mathscr{S}$ to take real values.  This generalization will have relevant implications in the dynamics of the qubit as it will be discussed later on.

The choice of an action $\mathscr{S}$ allows to extend Feynman's principle by associating to any history $w$ a complex number:
$$
\varphi(w) = C(w) \exp \frac{i}{\hbar} \mathscr{S}(w) \, ,
$$
where $C(w)$ is a real normalization constant and $\hbar$ is a constant that has the units of the action $\mathscr{S}$.    Note that $\varphi (w^{-1}) = C(w^{-1}) \exp -\frac{i}{\hbar} \overline{\mathscr{S}(w)}$.  Thus assuming that:
\begin{equation}\label{eq:C1}
C(\omega^{-1}) = C(\omega) \, ,
\end{equation}
we get the right Hermiticity property for the amplitudes: 
$$
\varphi (w^{-1}) = \overline{\varphi (w)} \, .
$$
 
Then, Feynman's use of the superposition principle can be extended easily to groupoids by stating that the probability amplitude $\langle a_f, t_f; a_i, t_i \rangle$ of observing the outcome $a_f$ at time $t_f$, after observing $a_i$ at time $t_i$ will be obtained as the linear superposition of the amplitudes of all histories $w \colon (a_i,t_i) \to (a_f,t_f)$, that is:
\begin{equation}\label{eq:feynman_histories}
\langle a_f, t_f; a_i, t_i \rangle =  \sum_{w \colon (a_i,t_i) \to (a_f,t_f)} C(w)  \exp \frac{i}{\hbar} \mathscr{S}(w) \, .
\end{equation}
Now it is clear that property (\ref{eq:S2}) together with (\ref{eq:C1}), implies the Hermiticity of the amplitudes, that is:
$$
\langle a_f, t_f; a_i, t_i \rangle = \overline{\langle a_i, t_i; a_f, t_f \rangle} \, .
$$

Given a history $w \colon (a_i,t_i) \to (a_f,t_f)$, choosing any  $t$, we can split the history $w$ as $w = w_2 \circ w_1$, with $w_1 \colon (a_t,t_i) \to (a,t)$, and $w_2 \colon (a,t) \to (a_f,t_f)$ (note that given any $w_1 \colon (a_t,t_i) \to (a,t)$, $w_2 = w \circ w_1^{-1}$ is fixed too).   In the previous formula $a$ is the target of $w(t)$ and $w_1 = w\mid_{[t_i,t]}$, and $w_2 = w\mid_{[t,t_f]}$, are the restrictions of the history $w$ to the subintervals $[t_i,t]$, and $[t,t_f]$, respectively.   Then we can write:
\begin{equation}\label{eq:decomposition}
\mathscr{G}_{(a_i,t_i}^{(a_f,t_f)} = \bigcup_{(a,t)} \mathscr{G}_{(a,t}^{(a_f,t_f)} \circ \mathscr{G}_{(a_i,t_i}^{(a,t)} \, ,
\end{equation}
meaning by that that each history $w \in \mathscr{G}_{(a_i,t_i}^{(a_f,t_f)}$ can be written as a composition of $w_1 \in \mathscr{G}_{(a_i,t_i}^{(a,t)}$ and $w_2 \in \mathscr{G}_{(a,t}^{(a_f,t_f)}$.

The decomposition property of the groupoid of histories above, \textit{cfr.} (\ref{eq:decomposition}), allows us to write the sum in  (\ref{eq:feynman_histories}) as:
\begin{eqnarray}
\langle a_f, t_f; a_i, t_i \rangle &=& \sum_{w \colon (a_i,t_i) \to (a_f,t_f)} C(w)  \exp \frac{i}{\hbar} \mathscr{S}(w) = \nonumber  \\ & =& \sum_{(a,t)} p(a,t)\left( \sum_{\tiny{\begin{array}{c} w_1\colon (a_i,t_i) \to (a,t) \\ w_2\colon (a,t) \to (a_f,t_f) \end{array}}}  C(w_2 \circ w_1)   \exp \frac{i}{\hbar} \mathscr{S}(w_2 \circ w_1)  \right) \label{eq:R1}
\end{eqnarray}
where we have introduced a probability measure $p(a)$ on $\Omega$ to compute the sum over all intermediate outcomes $a\in \Omega$. If we request that the normalization constants $C(w)$ satisfy:
\begin{equation}\label{eq:C2}
 C(w_1) C(w_2) = p(a,t)  C(w_1 \circ w_2 ) \, ,
\end{equation}
then, following the computation in (\ref{eq:R1}), we get:
\begin{eqnarray*}
\langle a_f, t_f; a_i, t_i \rangle &=& \sum_{(a,t)} \left( \sum_{\tiny{\begin{array}{c} w_1\colon (a_i,t_i) \to (a,t) \\ w_2\colon (a,t) \to (a_f,t_f) \end{array}}} C(w_2)  \exp \frac{i}{\hbar} \mathscr{S}(w_2 )  C(w_1)  \exp \frac{i}{\hbar} \mathscr{S}( w_1)  \right) \\ &=& \sum_{(a,t)} \sum_{w_2 \colon (a,t) \to (a_f,t_f)} \varphi (w_2) \sum_{w_1 \colon (a_i,t_i) \to (a,t)} \varphi (w_1) \, ,
\end{eqnarray*}
and we obtain the following reproducing property for the amplitudes $\langle a_f, t_f; a_i, t_i \rangle$,
\begin{equation}\label{eq:R2}
\langle a_f, t_f; a_i, t_i \rangle = \sum_{a\in \Omega}  \langle a_1, t_1; a, t \rangle \langle a, t; a_0, t_0 \rangle\, ,
\end{equation}
or, what is the same, writing it in integral notation:
$$
\langle a_f, t_f; a_i, t_i \rangle = \int   \langle a_f, t_f; a, t \rangle \langle a, t; a_i, t_i \rangle \, da \, .
$$
A natural choice for the normalization constants $C(w)$ satisfying conditions (\ref{eq:C1})-(\ref{eq:C2}), is given by:
\begin{equation}\label{eq:Cpp}
C(w) = \sqrt{p (s(w)) p (t(w))} \, .
\end{equation}

Continuing with discrete histories, it is obvious that the additive nature of the action, Eq. (\ref{eq:S1}), implies that there is a function $\ell \colon \Gamma \to \mathbb{C}$, such that if $w : \alpha_r, \ldots, \alpha_1$ is the sequence of steps defining a future-oriented history, $w \colon (a_i,t_i) \to (a_f, t_f)$, then:
$$
\mathscr{S}(w) = \sum_{k = 1}^r \ell (\alpha_k) \tau \, ,
$$
where $\tau$, the ``unit of time''  will be a real parameter such that $(t_1 - t_0) = N \tau$, with $N$ a positive integer.   Condition (\ref{eq:S2}) will be satisfied imposing that for $w^{-1}$, a past-oriented history, the inverse of the future-oriented history $w$, we get:
$$
\mathscr{S}(w^{-1}) = - \sum_{k = 1}^r \overline{\ell (\alpha_k)} \tau \, ,
$$
and, for a composite history $w = w_s \circ \cdots \circ w_1$, where $w_l$ is future or past-oriented, we will get:
$$
\mathscr{S} (w) = \sum_{l = 1}^s \mathscr{S}(w_l) = \sum_{l = 1}^s \epsilon (w_l)  \sum_{k_l = 1}^{r_l} \ell (\alpha_{k_l}) \tau \, , \qquad \sum_l r_l = N \, ,
$$
with $\epsilon (w_l) = \pm$, for $w_l$ being future or past-oriented respectively. 

It is also evident that in the case of continuous histories, the additivity property of $\mathscr{S}$ (together with the differentiability of the function $t \to \mathscr{S}(w(t))$), will imply the existence of a function $\ell \colon \Gamma \to \mathbb{C}$, such that;
\begin{equation}\label{eq:lagrangian}
\mathscr{S}(w) = \int_{t_0}^{t_1} \ell (w(t)) dt \, ,
\end{equation}
and condition (\ref{eq:S2}), will be satisfied provided that  $\ell (w^{-1}(t)) = \overline{\ell (w(t))}$.
Thus a Lagrangian for a quantum system described by the groupoid $\Gamma$ will be a function $\ell \colon \Gamma \to \mathbb{C}$, such that 
\begin{equation}\label{eq:I1}
\ell (\alpha) = \overline{\ell (\alpha^{-1})} \, .
\end{equation}
The function $\ell$ determined by the action $\mathscr{S}$, (\ref{eq:lagrangian}), will be called the Lagrangian of the theory.  We may also call it the groupoidal Lagrangian or the $q$-Lagrangian of the theory to emphasize that such Lagrangian is a purely quantum object, defined on the groupoid $\Gamma$ describing the quantum system and, in principle, is unrelated to any classical Lagrangian function.      Even more, the defining property (\ref{eq:lagrangian}) is telling us that $\ell$ defines an observable of the theory, that is, a self-adjoint element on the von Neumann algebra of the groupoid $\Gamma$.  In this sense $\ell$ is as in Schwinger's picture of quantum mechanics, an operator acting on some Hilbert space, while at the same time is a function on a set taking numerical values, so retaining some of its ``classical'' nature like in Feynman's description of quantum mechanical systems.    

Furthermore, in the simple situation of Feynman's paths, we identify paths $\gamma = \gamma(t)$, $\gamma \colon [t_0,t_1] \to Q$, with histories $w (t) = (\gamma(t), t_0)$.  Then, the inverse $w^{-1}$ of the history $w$ corresponds to the same path $\gamma (t)$ but with the opposite orientation in the interval $[t_0,t_1]$, that is, we replace $dt \mapsto -dt$.   In mechanical Lagrangian systems defined on the groupoid of pairs $P(Q)$ of some configuration space $Q$, the standard classical Lagrangian $L$ is defined not in the groupoid $P(Q) = Q \times Q$ of the theory, but in its infinitesimal counterpart $TQ$.     The evolution phase space $TQ$ is the Lie algebroid of the Lie groupoid $P(Q)$ ($Q$ is a smooth manifold) and there is a natural relation between both spaces (see \cite{Ci21a} for a detailed discussion on such relation).     Given a classical Lagrangian $L$ defined on $TQ$ we can associate to it a $q$-Lagrangian $\ell$ on the groupoid $P(Q)$ provided that we choose for any $\alpha$ in $\Gamma$ a history $w$ such that $w(t_1) = \alpha$.   More precisely, given the interval $[t_0,t_1]$ consider the natural projection $\pi \colon \mathscr{G} \to \Gamma$, $w \mapsto w(t_1) \in \Gamma$.  Then choose a cross section $\sigma \colon \Gamma \to \mathscr{G}$, such that $\pi \circ \sigma (\alpha) = \alpha$, for all $\alpha$.   Then, given $L$ a classical Lagragian function on $TQ$ and $\sigma$ a cross section of the map $\mathscr{G} \to Q\times Q$, we define:
$$
\ell (\alpha) = L (\dot{\sigma}(\alpha)(0)) \, .
$$

 Then it is clear that Lagrangians $\ell$ associated to classical mechanical Lagrangians of the form $L (q,\dot{q}) = \frac{1}{2} m \dot{q}^2 - V(q)$ satisfy the invariance property (\ref{eq:I1}).

After the previous discussion we are ready to consider the Lagrangian for the qubit system.  The most general self-adjoint function $\ell \colon A_2 \to \mathbb{C}$, that can be defined on the qubit grupoid $A_2$ is given by:
\begin{equation}\label{eq:L2}
\ell (1_+)=-V_+; \qquad \ell(1_-)=-V_-; \qquad \ell(\alpha)=\mu + i \delta; \qquad \ell(\alpha^{-1})=\mu - i \delta \, ,
\end{equation}
with $V_\pm$, $\mu$, $\delta$, real numbers.  The notation is chosen in accordance with the structure of classical mechanical Lagrangians of the form $L = K - V$, where $K$ is the kinetic energy of the system and $V$ a potential energy.    The value of $\ell$ at the units $1_\pm$ can be thought as a potential term while the value of $\ell$ on the transitions $\alpha$ and $\alpha^{-1}$ can be interpreted as a kinematical term.     The real part $\mu$ of $\ell (\alpha)$ would have the meaning of a kinetic energy and the imaginary part $\delta$ will represent a new kinematic contribution to the theory.   We will be back to this interpretation after the computation of the propagator of the theory in the coming section.


\subsection{Computing amplitudes in the groupoid of histories}\label{sec:feynman_path}

The explicit computation of the amplitude $\langle a_f, t_f; a_i, t_i \rangle$ using the extended Feynman's principle (\ref{eq:feynman_histories}), or, using an integral notation:
\begin{equation}\label{eq:P1}
\langle a_f, t_f; a_i, t_i \rangle = \int_{w \colon (a_i,t_i) \to (a_f,t_f)} \mathcal{D}w \,\, C(w) e^{\frac{i}{\hbar} \mathscr{S}(w)} \, ,
\end{equation}
is, in general, quite difficult.   However the algebraic structure of the space of histories allows to use a simple trick that simplifies notably the computation and that, in fact, would allow us to compute easily the amplitude for the qubit.   

The main idea was stated already at the end of Sect. \ref{sec:histories} and is based on the fact that any history can be written as the composition of a given reference history and a loop.   In fact, because the space of histories $\mathscr{G}$ of a groupoid is a groupoid itself, then any history $w \colon (a_i,t_i) \to (a_f,t_f)$ can be written as the composition of an ancillary or ``reference ''history $w_0 \colon (a_i,t_i) \to (a_f, t_f)$ and a loop $\sigma \colon (a_i,t_i) \to (a_i,t_i)$,  based at $(a_i,t_i)$.  Indeed, it suffices to show that $w$ can be written as $w = w_0 \circ (w_0^{-1} \circ w)$, then $\sigma = w_0^{-1}\circ w \colon (a_i,t_i) \to (a_i,t_i)$.    Notice that, similarly, we can write $w = \sigma' \circ w_0$, with $\sigma' \colon (a_f,t_f) \to (a_f,t_f)$ being a loop at $(a_f,t_f)$.   Thus given  $(a_i,t_i)$, $(a_f,t_f)$, fixing once for all an ancillary history $w_0 \colon (a_i,t_i) \to (a_f,t_f)$, the space of histories with origin $(a_i,t_i)$ and end $(a_f,t_f)$ can be identified with the isotropy group $\mathscr{G}(a_i,t_i)$ of $\mathscr{G}$ at $(a_i,t_i)$, that is, with the space of loops at $(a_i,t_i)$, by means of the map $w \in \mathscr{G} \mapsto \sigma = w_0^{-1}\circ w \in \mathscr{G}(a_i,t_i)$.  Then, we can write (\ref{eq:P1}) as:
\begin{eqnarray}
\langle a_f, t_f; a_i, t_i \rangle &=& \int_{w \colon (a_i,t_i) \to (a_f,t_f)} \mathcal{D}w \,\, C(w) e^{\frac{i}{\hbar} \mathscr{S}(w)} \nonumber \\ &=& \int_{\sigma \colon (a_i,t_i) \to (a_i,t_i)} \mathcal{D}\sigma \,\, \mathcal{J}_{w_0} (\sigma) C(w_0 \circ \sigma) e^{\frac{i}{\hbar} \mathscr{S}(w_0 \circ \sigma)} \nonumber \\ &=& 
C(w_0) e^{\frac{i}{\hbar} \mathscr{S}(w_0)} 
  \int_{\sigma \colon (a_i,t_i) \to (a_i,t_i)} \mathcal{D}\sigma \,\, \mathcal{J}_{w_0} (\sigma)  e^{\frac{i}{\hbar} \mathscr{S}(\sigma)} \label{eq:A3}\, ,
\end{eqnarray}
where $\mathcal{J}_{w_0}(\sigma)$ denotes the Jacobian of the transformation $\sigma \mapsto w$, and we have used the properties (\ref{eq:S1}) of the action and the fact that $C(w) = C(w_0)$.  Most important in the final expression (\ref{eq:A3}), is that the factor, in general a complex number:
\begin{equation}\label{eq:Gamma}
\Gamma(w_0) :=  \int_{\sigma \colon (a_i,t_i) \to (a_i,t_i)} \mathcal{D}\sigma \,\, \mathcal{J}_{w_0} (\sigma)  e^{\frac{i}{\hbar} \mathscr{S}(\sigma)}  \, ,
\end{equation}
is an integral over a group (the isotropy group of $\mathscr{G}$ at $(a_0,t_0)$) and depends solely on the choice of the ancilla $w_0$ (appart from the action $\mathscr{S}$ and the precise definition of $\mathcal{D}\sigma$, of course).   Then we obtain our main result in this section on the computation of amplitudes in the form of the simple formula:
\begin{equation}\label{eq:A_final}
\langle a_f, t_f; a_i, t_i \rangle = \Gamma (w_0) C(w_0) e^{\frac{i}{\hbar} \mathscr{S}(w_0)} \, ,
\end{equation}
for $w_0$ a reference history connecting $(a_i,t_i)$ to $(a_f,t_f)$.    

The procedure described above is reminiscent of the saddle point method for the (perturbative) computation of amplitudes.  There, the reference history is chosen as a classical solution of the Euler-Lagrange equations of the theory, and the computation proceeds by taking a perturbative expansion around it.   The method described here does not require the reference history $w_0$  to be a classical solution of the equations of motion, although this could be done once the appropriate Euler-Lagrange equations for the $q$-Lagrangian $\ell$ have been established.       An additional feature of formula (\ref{eq:A_final}) is that it is an integral over a group of a function satisfying a cocycle identity.  We will not insist here on the analysis of such integrals that will be discussed elsewhere.

Before embarking in the actual computation of the probability amplitudes for the qubit that will be done in the coming section, we can analyse further the structure of the coefficient $\Gamma (w_0)$ in (\ref{eq:A_final}).   The first thing to do is to understand better the nature of the Jacobian $\mathcal{J}(w)$ appearing in the definition of $\Gamma (w_0)$, Eq. (\ref{eq:Gamma}).    Following Feynman's insight, Eq. (\ref{eq:measure_paths}), the integral $\int \mathcal{D} w$ should be understood as an iterated integral $\int d \alpha_{N-1} d\alpha_{N-2} \cdots d\alpha_2 d \alpha_1$, on the intermediate steps of the history where, in the discrete situation we are considering, the elements ``$d\alpha_k$'', should be understood as appropriate weights for the steps of the history $w$.   Then, computing $\mathcal{D}(w_0 \circ w)$ will imply the computation of $\int d \beta_r d \beta_{r-1} \cdots d \beta_1d \alpha_{N-1} d\alpha_{N-2} \cdots d\alpha_2 d \alpha_1$, with $w_0 : \beta_1, \ldots, \beta_r$, fixed.  Then, the integrals $\int d \beta_r d \beta_{r-1} \cdots d \beta_1$ will be given by the weights of the steps of $\beta$.   Hence, $\mathcal{D}(w_0 \circ w)$ differs from $\mathcal{D}w$, in a factor that would depend just on the weights of the steps of the history $w_0$.  Denoting such factor as $\Delta (w_0)$, we get 
\begin{equation}\label{eq:delta}
\mathcal{D}(w_0 \circ w) = \Delta (w_0) \mathcal{D}w \, ,
\end{equation}
and the Jacobian $\mathcal{J}_{w_0}$ of the transformation is just $\Delta (w_0)$.    Then, because of the chain rule for Jacobians, we get:
$$
\Delta (w \circ w') = \Delta (w) \Delta(w') \, .
$$
Moreover $\Delta (w^{-1}) = \Delta(w)^{-1}$, because $\mathcal{D}\sigma = \mathcal{D }(w^{-1} \circ w) \circ \sigma = \Delta (w^{-1} \circ w) \mathcal{D}\sigma$, and then using (\ref{eq:delta}) we get the desired formula.   
The previous discussion shows us that 
$$
\Gamma (w) = \Delta (w) Z(a) \, ,
$$ 
where $w \colon (a,t) \to (b,t')$, and:
$$
Z(a) = \int_{\sigma \colon (a,t) \to (a,t)} \mathcal{D}\sigma \, e^{\frac{i}{\hbar} \mathscr{S}(\sigma)} \, ,
$$
that we will call the vertex integral at $a$.

The previous discussion shows that the expression (\ref{eq:A_final}) we got for the amplitude is well defined, that is, do not depend on the reference history  $w_0$ we choose to obtain it.   In fact, if $w_0' \colon (a_i,t_i) \to (a_f,t_f)$, where another reference history, then there is a loop $\sigma \colon (a_i,t_i) \to (a_i,t_i)$ such that $w_0' = w_0 \circ \sigma$ ($\sigma = w_0^{-1} \circ w_0'$).  Then,
\begin{eqnarray*}
\Gamma (w_0') C(w_0') e^{\frac{i}{\hbar} \mathscr{S}(w_0')} &=& \Gamma (w_0\circ \sigma)  C(w_0) e^{\frac{i}{\hbar} \mathscr{S}(w_0\circ \sigma)} \\ &=& 
\Delta (w_0\circ \sigma) Z(a) C(w_0) e^{\frac{i}{\hbar} \mathscr{S}(w_0)} e^{\frac{i}{\hbar} \mathscr{S}(\sigma)}  \\ &=&
\Delta (w_0)  Z(a) C(w_0) e^{\frac{i}{\hbar} \mathscr{S}(w_0)}  \Delta( \sigma)e^{\frac{i}{\hbar} \mathscr{S}(\sigma)} \\ &=& \Gamma (w_0)  C(w_0) e^{\frac{i}{\hbar} \mathscr{S}(w_0)}\Delta( \sigma)e^{\frac{i}{\hbar} \mathscr{S}(\sigma)}  \, ,
\end{eqnarray*}
then, defining $\Delta(\sigma) = e^{-\frac{i}{\hbar} \mathscr{S}(\sigma)}$, we get $\Gamma (w_0') C(w_0') e^{\frac{i}{\hbar} \mathscr{S}(w_0')} = \Gamma (w_0) C(w_0) e^{\frac{i}{\hbar} \mathscr{S}(w_0)}$ and the amplitude $\langle a_f,t_f;a_i,t_i \rangle$ given by (\ref{eq:A_final}) is well-defined.

\subsection{A path integral computation of the qubit probability amplitudes}\label{sec:qubit_amplitude}

We are now ready to compute the amplitudes of the qubit whose dynamics is given by the $q$-Lagrangian $\ell$ in (\ref{eq:L2}). Before doing that we will obtain a few more formulas that will be quite useful in our discussion.    We will do it in the simple context of groupoids with a finite number of outputs, like the qubit groupoid $A_2$, and we will use discrete histories on them.  Thus, we will assume that the space of outputs $\Omega$ of the groupoid $\Gamma$ is finite and we label its elements as $a_k$, $k = 1, \ldots, n$.   Then, fixing the time interval $[t_i,t_f]$ and the number of steps $N$, $\tau = (t_f - t_i)/N$, we will consider the amplitudes $\langle a_f, t_f; a_i, t_i \rangle$ as the entries of a $n\times n$ matrix:
$$
U (t_f,t_i) := (U_{lk}(t_f, t_i)) = (\langle a_l, t_f; a_k, t_i \rangle ) \, , \qquad k,l = 1, \ldots, n \, .
$$
With this notation the entries of the unit time matrix $U(t_i + \tau, t_i)$ are given by $U_{lk}(t_i + \tau, t_i) = \langle a_l, t_i + \tau; a_k, t_i \rangle $.
In what follows we will just consider autonomous systems, that is systems such that:
$$
U(t_f, t_i ) = U(t_f + T, u_i + T) \, ,
$$
for all $T\in \mathbb{R}$.    In such case, we will just denote by $U_\tau$ the unit time operator, that is $(U_\tau)_{lk} = \langle a_l, t_i + \tau; a_k, t_i \rangle$.      Notice that Eq. (\ref{eq:R2}) can be written with the new notation as:
$$
U_{lk}(t_f,t_i) = \sum_{j= 1}^n  U_{lj}(t_f,t)   U_{jk}(t, t_i) \, ,
$$
for an arbitrary $t \in [t_i,t_f]$.  The previous equation can be written as:
$$
U(t_f, t_i) = U(t_f, t) U(t, t_i) \, .
$$
 Note that, in particular, $U(t_i + 2 \tau, t_i) = U_\tau U_\tau$, and, in general, we get:
$$
U(t_f , t_i) =  U_\tau^N \, .
$$
 The $n\times n$ matrix $U_\tau$ will be called the unit time discrete propagator of the theory.     The propagator $\mathbf{U}(t_f,t_i)$ of the theory will be obtained as the continuous  limit of $U_\tau^N$, that is:
 $$
 \mathbf{U}(t_f, t_i) = \lim_{\stackrel{N \to \infty}{\tau \to 0}} U_\tau^N \, .
 $$
In general such limit will be difficult to compute explicitly and we will use the discrete propagator $U_\tau$ to describe the evolution of the system.   Certainly, if $|\Psi_i \rangle \in \mathcal{H}_n$ is a pure state of the system at time $t_i$ (recall \S \ref{sec:states}), then the evolved state at time $t_f$ will be given by:
$$
| \Psi_f \rangle = U(t_f, t_i) | \Psi_i \rangle = U_\tau^N | \Psi_i \rangle \, .
$$\\

\medskip

The space of outcomes of the qubit has only to elements $\{ + , - \}$, thus in the particular instance of the qubit groupoid $A_2$, the unit time propagator will be a $2 \times 2$ matrix $U_\tau$ of the form:
\begin{equation}\label{eq:Uqubit}
U_\tau = \left( \begin{array}{cc}  U_{--} & U_{-+} \\ U_{+-} & U_{++} \end{array}\right) \, ,
\end{equation}
whose entries, according to (\ref{eq:A_final}), will be given by :
\begin{eqnarray}
U_{--} &=& \Gamma_{--} C(w_{--}) e^{\frac{i}{\hbar} \ell (w_{--}) \tau} \, , \qquad U_{-+} = \Gamma_{-+} C(w_{-+}) e^{\frac{i}{\hbar} \ell (w_{-+}) \tau}, \\
U_{+-} &=& \Gamma_{+-} C(w_{+-}) e^{\frac{i}{\hbar} \ell (w_{+-}) \tau} \, , \qquad U_{++} = \Gamma_{++} C(w_{++}) e^{\frac{i}{\hbar} \ell (w_{++}) \tau}  \, ,
\end{eqnarray}
where $w_{--}$ is the reference history $w_{--} \colon (-, t_i) \to (-,t_i + \tau)$, and similarly for $w_{-+} \colon  (-, t_i) \to (+,t_i + \tau)$, $w_{-+} \colon  (+, t_i) \to (-,t_i + \tau)$, and $w_{++} \colon  (+, t_i) \to (+,t_i + \tau)$.     Because in the case of the qubit groupoid $A_2$, there is only one transition among any two outcomes, there is no ambiguity in the choice of the reference histories $w_{\pm,\pm}$ and they are given as:
$$
w_{--} = 1_- \, , \quad w_{-+} = \alpha \, , \quad w_{+-} = \alpha^{-1} \, , \quad w_{++} = 1_+ \, .
$$
Using (\ref{eq:Cpp}) for the normalization constants $C(w)$, and denoting by $p_\pm = p(\pm)$, the entries $U_{\pm \pm}$ of the propagator become:
\begin{eqnarray}
U_{--}  &=& \Gamma_{--}  \sqrt{p_-p_-}e^{\frac{i}{\hbar}l(1_-)\tau }=\Gamma_{--}  \sqrt{p_-p_-}e^{-\frac{i}{\hbar}V_-\tau } \label{eq:Uentries1} \\
U_{-+} &=& \Gamma_{-+}  \sqrt{p_-p_+}e^{\frac{i}{\hbar}l(\alpha^{-1})\tau }=\Gamma_{-+}  \sqrt{p_-p_+}e^{\frac{i}{\hbar}(\mu - i \delta)\tau } \label{eq:Uentries2} \\
U_{++} &=& \Gamma_{++} \sqrt{p_+p_+}e^{\frac{i}{\hbar}l(1_+)\tau }=\Gamma_{++}  \sqrt{p_+p_+}e^{-\frac{i}{\hbar}V_+\tau }\label{eq:Uentries3} \\
U_{+-} &=& \Gamma_{+-}  \sqrt{p_+p_-}e^{\frac{i}{\hbar}l(\alpha)\tau }= \Gamma_{+-}  \sqrt{p_+p_-}e^{\frac{i}{\hbar}(\mu + i \delta)\tau} \label{eq:Uentries4} 
\end{eqnarray}
Let us recall that the coefficients $\Gamma_{\pm \pm}$ in the previous formulas denote the integrals (\ref{eq:Gamma}) on the spaces of loops $\sigma$ based at the points $(\pm, t_i)$.   We will not try to compute directly the coefficients $\Gamma_{\pm\pm}$ but we will rather proceed in an indirect way by imposing that the propagator $U_\tau$ must be unitary. 

The probabilities $p_+, p_-$, represent an intrinsic bias of the system, that is, the outcomes $+,-$ are biased according to the probability distribution $(p_+, p_-)$.   As it was pointed out at the end of Sect. \ref{sec:observables} such bias is instrumental in the construction of infinite spin chains.   In our simple situation we will denote $0 \leq  p_+ \leq 1/2$ by $p$ and $p_- = 1 - p$.     The particular instance of $p = 1/2 = 1-p$ will be called the basic or uniform qubit.

Then, imposing unitarity to the matrix $U_\tau$, cfr. (\ref{eq:Uqubit}), that we write explicitly below:
\begin{equation}\label{Ugrupoides}
U_\tau = \left( \begin{array}{ll}
  \Gamma_{--} p_- e^{-\frac{i\tau}{\hbar}V_- } & \Gamma_{-+} \sqrt{p_-p_+}e^{\frac{\tau}{\hbar}(-\delta + i \mu) }\\
  \Gamma_{+-}\sqrt{p_+p_-}e^{\frac{\tau}{\hbar}(\delta + i \mu)} & \Gamma_{++} p_+ e^{-\frac{i\tau}{\hbar}V_+} 
\end{array}\right)
\end{equation}
we obtain the following relations:
\begin{align}
    &\label{ecuacion1.1} |\Gamma_{--}|^2 p_-^2 + |\Gamma_{-+}|^2 p_+p_- e^{-\frac{2\delta \tau}{\hbar}} = 1 \\
    &\label{ecuacion1.2} \overline{\Gamma_{--}}\Gamma_{+-} p_-\sqrt{p_+p_-}e^{\frac{\tau}{\hbar}(\delta-i\mu-iV_-)} + \overline{\Gamma_{-+}} \Gamma_{++} p_+\sqrt{p_+p_-}e^{\frac{\tau}{\hbar}(-\delta+i\mu+iV_+)}=0 \\
    &\label{ecuacion1.3} \overline{\Gamma_{+-}} \Gamma_{--}p_-\sqrt{p_+p_-}e^{\frac{\tau}{\hbar}(\delta+i\mu+iV_-)}+ \overline{\Gamma_{++}}\Gamma_{-+} p_+\sqrt{p_+p_-}e^{\frac{-\tau}{\hbar}(\delta+i\mu+iV_+)}=0 \\
    &\label{ecuacion1.4} |\Gamma_{++}|^2 p_+^2 + |\Gamma_{+-}|^2 p_ + p_-e^{\frac{2\delta \tau}{\hbar}}=1
\end{align}
(from $U_\tau U_\tau^\dagger = I$), or:
\begin{align}
    &\label{ecuacion2.1} |\Gamma_{--}|^2 p_-^2 + |\Gamma_{+-}|^2 p_+ p_- e^{\frac{2\delta \tau}{\hbar}}=1\\
    &\label{ecuacion2.2}\overline{\Gamma_{--}} \Gamma_{-+} p_-\sqrt{p_+p_-}e^{\frac{\tau}{\hbar}(-\delta+i\mu+iV_-)}+\overline{\Gamma_{+-}} \Gamma_{++}p_+\sqrt{p_+p_-}e^{\frac{\tau}{\hbar}(\delta-i\mu-iV_+)}=0 \\
    &\label{ecuacion2.3} \overline{\Gamma_{-+}} \Gamma_{--}p_-\sqrt{p_+p_-}e^{\frac{-\tau}{\hbar}(\delta+i\mu+iV_-)}+ \overline{\Gamma_{++}} \Gamma_{+-}p_+\sqrt{p_+p_-}e^{\frac{\tau}{\hbar}(\delta+i\mu+iV_+)}=0\\
    &\label{ecuacion2.4} |\Gamma_{++}|^2p_+^2 + |\Gamma_{-+}|^2p_+p_-e^{\frac{-2\delta \tau}{\hbar}}=1 \, ,
\end{align}
from $U_\tau^\dagger U_\tau = I$.   From equations \eqref{ecuacion1.1} and \eqref{ecuacion2.1} the resulting relationship is:
\begin{equation}\label{relacion1}
\frac{|\Gamma_{-+}|}{|\Gamma_{+-}|} = e^{\frac{2\tau\delta}{\hbar}} \, ,
\end{equation}
i.e., we get the following relation between the factors $\Gamma_{-+}$ and $\Gamma_{+-}$:
$$
\Gamma_{-+} = \Gamma_{+-} e^{\frac{2\delta \tau}{\hbar}} e^{\frac{i}{\hbar}\Lambda} \, ,
$$
where $\Lambda$ is a phase factor,
whereas from equations \eqref{ecuacion1.1} and \eqref{ecuacion2.4} one finds:
$$
|\Gamma_{--}|^2 p_-^2 = |\Gamma_{++}|^2p_+^2 \, ,
$$
or, equivalently:
\begin{equation}\label{relacion2}
\frac{|\Gamma_{--}|}{|\Gamma_{++}|}= \frac{p_+}{p_-}=  \frac{1-p}{p}  \, ,
\end{equation}
and
$$
\Gamma_{--} p_- = \Gamma_{++} p_+ e^{\frac{i}{\hbar}\Sigma} \, , 
$$
with $\Sigma$ another phase factor.
Inserting the relations (\ref{relacion1})-(\ref{relacion2}) into equations \eqref{ecuacion2.2} and \eqref{ecuacion2.3}, we get:
$$
 \frac{p_+}{p_-}\Gamma_{++}\overline{\Gamma_{+-}} e^{\frac{2\tau\delta}{\hbar}} e^{-\frac{i}{\hbar}(\Sigma+\Lambda)}p_- \sqrt{p_+p_-}e^{\frac{\tau}{\hbar}(-\delta+i\mu+iV_-)} +\overline{ \Gamma_{+-}} \Gamma_{++} p_+\sqrt{p_+p_-}e^{\frac{\tau}{\hbar}(\delta-i\mu-iV_+)} = 0 \, ,
$$
namely, 
\begin{equation}\label{eq:global}
e^{i \left(\mu+\frac{V_++V_-}{2}\right) \frac{2\tau}{\hbar} } =  e^{i\left(\frac{\Sigma+\Lambda}{\hbar} + \pi\right)} \, .
\end{equation}
Thus, for instance, if both $\Gamma_{--}$ and $\Gamma_{++}$, were real, then $\Lambda$, and $\Sigma$ will be either $0$, or $\hbar \pi$, and then:
\begin{equation}\label{eq:quantizing}
\frac{2\tau }{\hbar}\left(\mu+\frac{V_++V_-}{2}\right) = n \pi \, , \qquad n \in \mathbb{Z} \, ,
\end{equation}
that could be restated as:
\begin{equation}\label{cuantizacionlagr}
 \mu+\frac{V_++V_-}{2} = ( \pm \frac{1}{2}, \pm \frac{3}{2} , \ldots ) \frac{\pi \hbar}{\tau} ,
\end{equation}
that shows that the ``energy'' $\mu + \bar{V}$, $\bar{V} = (V_+ +V_-) /2$, can only take discrete values in units of the Planck constant divided by $\tau$.   

In conclusion, imposing the unitarity condition on the propagator not only yields relations between the factors $\Gamma_{\pm \pm}$ (\textit{cfr.} Eqs. \eqref{relacion1}-\eqref{relacion2}), but also a surprising restriction for the parameters $\mu, V_\pm$, defining the Lagrangian of the system. Not all Lagrangians for the system are possible, its value is quantized, \textit{cfr.} \eqref{eq:global}.   

As it turns out, the general expression for the discrete propagator $U_\tau$ becomes:
$$
U_\tau = \left( \begin{array}{cc}
  \Gamma_{--} p_- e^{-\frac{i\tau}{\hbar}V_- } & \Gamma_{-+} \sqrt{p_-p_+}e^{\frac{\tau}{\hbar}(-\delta + i \mu) }\\
  \Gamma_{-+}\sqrt{p_+p_-}e^{\frac{\tau}{\hbar}(-\delta + i \mu)} e^{-\frac{i}{\hbar}\Lambda}  &   \Gamma_{--} p_- e^{-\frac{i\tau}{\hbar}V_- } e^{-\frac{i}{\hbar}\Sigma}
  \end{array} \right)  = \left( \begin{array}{cc}
  A & B \\
 B e^{-\frac{i}{\hbar}\Lambda}  &  A e^{-\frac{i}{\hbar}\Sigma}
  \end{array} \right)  \,.
$$
with $A = \Gamma_{--} p_- e^{-\frac{i\tau}{\hbar}V_- }$, and $B =  \Gamma_{-+} \sqrt{p_-p_+}e^{\frac{\tau}{\hbar}(-\delta + i \mu) }$.   It is clear from the form of the matrix $U_\tau$ that in general is not possible to provide a simple expression for the continuous limit $\mathbf{U}(t_f - t_i)$.   However the structure of the propagator get simpler in some particular cases of interest.   If we consider the ``free'' unbiased  qubit, that is, $V_\pm$, and $p_+ = p_- = 1/2$, then it is reasonable to assume that the integrals defining $\Gamma_{-+}$ and $\Gamma_{+-}$ on one side, and $\Gamma_{--}$, $\Gamma_{++}$ on the other, Eq. (\ref{eq:Gamma}), are equal and, consequently, $\Gamma_{-+} = \Gamma_{+-} = \Gamma'$, $\Gamma_{--} = \Gamma_{++} = \Gamma$,  $\Lambda = \Sigma =  0$.  Under such circumstances we get:
$$
U_\tau = \frac{1}{2}  \left( \begin{array}{cc}
  \Gamma & \Gamma' e^{i \frac{\mu \tau}{\hbar}}\\
 \Gamma' e^{i \frac{\mu \tau}{\hbar}} &  \Gamma
  \end{array} \right)  \,.
$$ 
The eigenvalues of the matrix $U_\tau$ are given by: $\lambda_{\pm} = \Gamma \pm \Gamma' e^{i\frac{\mu \tau}{\hbar}}$, and the eigenvalues of the continuous propagator $\mathbf{U}(T)$ behave as $C_\pm e^{i\mu T}$.    Note that, in this case, the ``mass'' term $\mu$ cannot take arbitrary values and Eqs. (\ref{eq:global})-(\ref{eq:quantizing}) imply that it must satisfy the quantization condition $\mu\tau = n\pi \hbar$, $n$ an integer.   

As indicated above, another interesting situation arises when $\Gamma$, $\Gamma'$ are both real, in which case, Eqs. (\ref{relacion1})-(\ref{relacion2}), become:
$$
\Gamma_{-+} = \pm \Gamma_{+-} e^{\frac{2\delta \tau}{\hbar}} \, , \qquad  \Gamma_{--} p_- = \pm  \Gamma_{++} p_+  \, .
$$
Moreover, if one considers the free case in which $V_+=V_-=0$, i.e., the ``free'' case, then we get for $\mu$:
$$
\frac{\tau}{\hbar}\mu=\frac{n}{2} \pi \, , \qquad n\in \mathbb{Z} \, ,
$$
and the propagator $U_\tau$, takes now the simple form:
\begin{equation}\label{eq:Upm}
U_\tau = \left( \begin{array}{cc}
  A & B \\
 \pm B   &  \pm A 
  \end{array} \right)  \,.
\end{equation}
with $A = \Gamma_{--}p_-$, and $B = \Gamma_{-+}\sqrt{p_+p_-} e^{(i\mu - \delta) \frac{\tau}{\hbar}}$.   
There are four cases for the spectrum of $U_\tau$ depending on the configuration of signs $+,-$ in the previous formula.  So we get:  (I) $\lambda_{\pm}^{(I)} = A\pm B$ (signs $+,+$ in (\ref{eq:Upm})); (II)  $\lambda_{\pm}^{(II)} = A\pm iB$ (signs $-,+$); (III)  $\lambda_{\pm}^{(III)} = \pm \sqrt{A^2 +  B^2}$ (signs $+,-$);  (IV) $\lambda_{\pm}^{(IV)} = \pm \sqrt{A^2 - B^2}$ (signs $-,-$).  

The details of the dynamics associated to each one of these situations as well as to other examples and situations will be discussed elsewhere.



\section{Discussion and conclusions}\label{sec:conclusions}

The Lagrangian representation and the corresponding extension of Feynman's path integral description for quantum system lacking a classical description, that is, quantum systems for which there are not a well-defined classical configuration space and classical Lagrangian $L$, has been presented.   The main idea is to describe such quantum systems by means of groupoids, following and extending Schiwnger's picture of Quantum Mechanics, and introduce a novel notion of Lagrangian as a function $\ell$ defined on the groupoid itself which is self-adjoint as an element of its corresponding algebra. Such function has been called the  $q$-Lagrangain of the theory.   In such a way the dynamics of the qubit is computed for its most general Lagrangian $\ell$ and some general relations for the propagator of the theory are obtained that, in some particular instances, lead to striking quantization conditions for its coefficients.   

The path integral formulation is obtained by exploiting the structure of the groupoid describing the quantum sytem and its corresponding spaces of histories (notion that extend Feynman's paths to this setting).  Even if a number of relevant properties for the path integral are obtained it would be necessary to conduct a more detailed analysis of it.   Such analysis will be conducted together with other relevant examples, like the $d$bit, $d > 2$, or infinite chains of spins (see the groupoidal treatment of such systems in \cite{Ci23}).

In particular, it is worth to point out that a general formula that allows to compute the probability amplitudes using arbitrary reference histories, has been obtained. However such expressions can be improved notably by using as reference histories solutions of the corresponding classical Euler-Lagrange equations for the $q$-Lagrangian $\ell$ on the groupoid (see, for instance, \cite{Ma06,Co06} for a detailed analysis of such equations).

Another relevant aspect of the theory is that it allows to compute approximations to quantum systems possessing a well-defined classical corresondence by introducing a coarse-graining approach.   In fact, decomposing the configuration space $Q$ into a family of subsets $A_1, \ldots, A_M$, we can deine a ñprojection form the groupoid of pairs $P(Q)=$ to the groupoid of classes of equivalent points and, averaging the original Lagrangian on such subsets, we will obtain a Lagrangian defined on a finite groupoid that can be used to compute discrete propagators approximating the original one.

A comparison with the standard Hamiltonian description will be conducted in subsequent works together with the study of the implications of the quantization conditions found for the paranmeters $\mu$, $V_\pm$ of the Lagrangian of the theory.
\backmatter

\bmhead{Acknowledgments}
The authors acknowledge financial support from the Spanish Ministry of Economy and Competitiveness, through the Severo Ochoa Programme for Centres of Excellence in RD (SEV-2015/0554), and the MINECO research project  PID2020-117477GB-I00.



\end{document}